\documentclass[final]{article}
\usepackage{graphicx,multirow}
\usepackage{cite}
\usepackage{url}
\usepackage{booktabs}
\usepackage{latexsym}
\usepackage{color}
\usepackage{colortbl,hhline}
\usepackage{fullpage}

\input{Qcircuit.tex}
\begin{document}

\title{\large \bf BLOCK-BASED QUANTUM-LOGIC SYNTHESIS}
\author{\normalsize Mehdi Saeedi, Mona Arabzadeh, Morteza Saheb Zamani, Mehdi Sedighi\\
\normalsize Quantum Design Automation Lab\\
\normalsize Department of Computer Engineering and Information Technology\\
\normalsize Amirkabir University of Technology\\
\normalsize Tehran, Iran \\
\normalsize \texttt{\{msaeedi, m.arabzadeh, szamani, msedighi\}@aut.ac.ir}
 }
 \date{The first two authors contributed equally to this work.}
\maketitle

\newcounter {TCounter}
\newcounter {LCounter}
\newcounter {ECounter}
\newcounter {CCounter}
\newcounter {DCounter}
\newcounter {FCounter}
\newcounter {PCounter}

\newtheorem{prop}[PCounter]{\textbf{Proposition}}

\newcommand{\G}[1]{\textcolor{gray}{{#1}}}

\begin{abstract}
In this paper, the problem of constructing an efficient quantum circuit for the implementation of an arbitrary quantum computation is
addressed.
To this end,
a basic block based on the cosine-sine decomposition method is suggested which contains $l$ qubits. In addition, a previously proposed quantum-logic synthesis method based on  quantum
Shannon decomposition is recursively applied to reach unitary gates over $l$ qubits. Then, the basic block is used and some optimizations are applied to remove redundant gates. It is shown that the exact value of $l$ affects the number of one-qubit and CNOT gates in the proposed method. In comparison to the previous synthesis methods, the value of $l$ is examined consequently to improve either the number of CNOT gates or the total number of gates.
The proposed approach is further analyzed by considering the nearest neighbor limitation.
According to our evaluation, the number of CNOT gates is increased by at most a factor of $\frac{5}{3}$ if the nearest neighbor interaction is applied.
\end{abstract}

\section{Introduction} \label{sec:intro}

The promise of exponential speed up of quantum algorithms \cite{Elliott, shor-1997-26, Grover} running on quantum computers has intensified the attempts for using quantum algorithms in real world problems. Besides that, the ability of quantum computation in simulating quantum-mechanical effects (QMEs) further increases its attraction. Classical computation has inherent limitations to simulate QMEs \cite{Feynman82}.
Due to the ability of quantum computation in solving several applications efficiently \cite{Mosca},
numerous efforts have been made to study various aspects of quantum computation and a set of quantum technologies has been proposed to build a scalable quantum computer \cite{Ross2008}.

Aside from different technological characterizations, quantum computations provided by all of the proposed quantum technologies can be described in terms of linear operators (i.e., matrices) in Hilbert space. In this sense, any quantum computation involves evolution of an initial quantum state under a series of elementary unitary transformations \cite{Deutsch89, DiVincenzo95, Barenco95, Lloyd95}.
It has been shown that any unitary transformation can be exactly realized if the set of single qubit operations plus CNOT are allowed as the elementary gates \cite{Barenco95, Nielsen00}.
As CNOT needs interactions between two qubits which are usually much weaker than the interactions between a single qubit and classical control fields, producing fewer CNOT gates is more favorable generally. At the same time, reducing the number of one-qubit gates to improve the total synthesis cost is an active research field.

Synthesis of an arbitrary unitary matrix from a universal set of gates including one-qubit operations and CNOTs has a rich history. Barenco et al. in 1995 \cite{Barenco95} showed that the number of CNOT gates required to implement an arbitrary unitary matrix over $n$ qubits was $O(n^34^n)$.
In \cite{George2001}, it was shown that applying the QR decomposition of matrix algebra could lead to the same result.
In 1995, Knill reduced the number of CNOT gates to $O(n4^n)$ \cite{knill95}. In 2004, Shende et al. found the highest known lower bound on the number of CNOT gates as $\lceil \frac{1}{4} (4^n - 3n - 1) \rceil$ \cite{shende-2004}. The process of reducing the number of CNOT gates was continued by Vartiainen et al. in \cite{vartiainen-2004} which led to $O(4^n)$ CNOT gates. Later, a decomposition method based on the cosine-sine matrix decomposition, called CSD, was proposed which produced $4^n - 2^{n+1}$ CNOT gates \cite{Mottonen-2004}.
In \cite{Bergholm:2004} uniformly controlled one-qubit gates were used by Bergholm et al. in CSD method which resulted in $\frac{1}{2} 4^n  - \frac{1}{2} 2^n  - 2$ CNOT gates and $\frac{1}{2}4^n  + \frac{1}{2} 2^n  - n - 1$ one-qubit gates.
Another quantum circuit decomposition, called NQ decomposition, based on CS decomposition was introduced by Shende et al. \cite{Shende05} which leads to $\frac{1}{2} 4^n - \frac{3}{2} 2^n + 1$ CNOT gates. Next, M\"{o}tt\"{o}nen and Vartiainen in \cite{Mottonen05} reduced the number of CNOT gates to $\frac{23}{48} 4^n - \frac{3}{2} 2^n + \frac{4}{3}$ by improving the results of \cite{Shende05}. By using the idea of quantum Shannon decomposition (QSD) and some circuit identities, the authors of \cite{Shende06} reached the same number of CNOT gates.
{In \cite{Nakajima:2006}, an algorithm was proposed that translated a unitary matrix into a quantum circuit according to the KAK decomposition in Lie group theory. Their results showed that any matrix decomposition corresponding to type-AIII KAK decompositions could be derived according to the given Cartan involution by applying the method of \cite{Nakajima:2006}. Most recently, Drury and Love placed the QSD into a Lie algebraic context. Their results showed that QSD is an alternating series of Cartan decompositions \cite{Drury08}.}

Besides the significant attempts made in improving the worst-case number of CNOT gates and one-qubit gates required to implement an arbitrary unitary matrix, efforts have been made to improve the number of one-qubit gates and CNOTs for specific matrices with applications in the synthesis of an arbitrary matrix. It has been shown that three CNOT gates are necessary and sufficient to implement an arbitrary 2-qubit unitary matrix in the worst case \cite{Shende-1-2004}, \cite{vidal-2004-69}, \cite{vatan-2004-69}. In addition, seven one-qubit gates are reported to be sufficient by three CNOT gates for implementing an arbitrary 2-qubit unitary matrix  \cite{Shende-1-2004}.
In \cite{Zhang04}, a new quantum gate, $B$, was introduced and applied to synthesize a generic two-qubit operation. It has been shown that two B gates with at most six single-qubit gates implement any arbitrary two-qubit quantum operation \cite{Zhang04}.
 On the other hand, a decomposition algorithm which leads to 40 CNOT gates was proposed in \cite{vatan-2004} for an arbitrary quantum circuit with 3 qubits. Currently, it appears that the best known quantum circuit for a 3-qubit matrix requires 20 CNOTs \cite{Shende06} whereas the lower bound is 14 \cite{shende-2004}. The synthesis of an $n$-qubit Toffoli gate was studied in \cite{Shende09} where the authors showed that an $n$-qubit Toffoli gate requires at least $2n$ CNOT gates and for $n=3$, their CNOT-cost (i.e., 6) is optimal. On different directions, synthesis of a diagonal matrix and optimal realization of controlled unitary gates were studied in \cite{bullock-2004-4} and \cite{song-2003-3}, respectively. {The problem of preparing an arbitrary quantum state starting from a given state was discussed in some papers \cite{Shende06, Bergholm:2004, Mottonen-2005}}.

In this paper, the synthesis of an arbitrary unitary matrix over $n$ qubits is addressed. To this end, the QSD method is used to synthesize a given unitary transformation first. Then, by determining an appropriate decomposition level, unstructured unitary matrices in the resulted decomposed circuit are replaced with blocks synthesized by CSD method equipped by some optimizations.
We show that the decomposition level can be selected to trade off the number of CNOT gates against one-qubit gates in the proposed decomposition method.

The remainder of the paper is organized as follows. In Section \ref{sec:basic_concepts} basic concepts are explained. The proposed synthesis approach is introduced in Section \ref{sec:our_method} followed by considering the number of CNOT and one-qubit gates resulted from the proposed approach in Section \ref{sec:results}. The nearest neighbor implementation of our synthesis method is evaluated in Section \ref{sec:NN} and finally, Section \ref{sec:conclusion} concludes the paper.

\section{Preliminaries} \label{sec:basic_concepts}
In this section, the basic concepts and notations used in the rest of the paper are introduced and explained.

\subsection{Quantum Bits}
A quantum bit, qubit in short, like its classical counterpart can be realized by a physical system such as a photon. In this paper, we treat a qubit as a mathematical object which represents a quantum state with two basic states $|0\rangle$ and $|1\rangle$. A qubit can get any linear combination of its basic states, called superposition, as shown in (\ref{Eq:dirac}) where $\alpha$ and $\beta$ are complex numbers.
\begin{equation}\label{Eq:dirac}
|\psi\rangle = \alpha|0\rangle+\beta|1\rangle
\end{equation}

Although a qubit can get any linear combination of its basic states, when a qubit is measured, its state collapses into the basis $|0\rangle$ and $|1\rangle$ with the probability of $|\alpha|^2$ and $|\beta|^2$, respectively and we have $|\alpha|^2+|\beta|^2= 1$. It is also common to denote the state of a single qubit by a $2 \times 1$ vector as $\left[ {\alpha \,\beta } \right]^T$ in Hilbert space $H$ where superscript $T$ stands for the transpose of a vector. A quantum system which contains $n$ qubits is often called a quantum register of size $n$ in the context of quantum computation. More precisely, an $n$-qubit quantum register can be described by an element $|\psi \rangle = |\psi_1 \rangle \otimes |\psi_2 \rangle \otimes \ldots \otimes |\psi_n \rangle$ in the tensor product Hilbert space $H = H_1 \otimes H_2 \otimes \dots \otimes H_n$.

\subsection{Quantum Gates and Circuits}
An $n$-qubit quantum gate is a device which performs a specific $2^n\times2^n$ unitary operation on selected $n$ qubits in a specific period of time. A matrix $U$ is unitary if $UU^\dag=I$ where $U^\dag$ is the conjugate transpose of $U$ and $I$ is the identity matrix. An arbitrary unitary gate over $n$ qubits with a generic $2^n\times2^n$ matrix is represented as $U(2^n)$ in this paper. Previously, various quantum gates with different functionalities have been introduced. For example, the $\theta$ rotation gates ($0 \leq \theta \leq 2\pi$) around the $x$, $y$ and $z$ axes acting on one qubit are defined as:\\\\
\vspace{10pt}
     $R_x (\theta ) = \left( {\begin{array}{*{20}c}
       {\cos \frac{\theta }{2}} & {i\sin \frac{\theta }{2}}  \\
       {i\sin \frac{\theta }{2}} & {\cos \frac{\theta }{2}}  \\
    \end{array}} \right)$ ,
     $R_y (\theta ) = \left( {\begin{array}{*{20}c}
       {\cos {\textstyle{\theta  \over 2}}} & {\sin {\textstyle{\theta  \over 2}}}  \\
       { - \sin {\textstyle{\theta  \over 2}}} & {\cos {\textstyle{\theta  \over 2}}}  \\
    \end{array}} \right)$,
    $R_z (\theta) = \left( {\begin{array}{*{20}c}
   {e{\textstyle{{ - i\theta } \over 2}}} & 0  \\
   0 & {e{\textstyle{{i\theta } \over 2}}}  \\
    \end{array}} \right)$

A square matrix is called diagonal if the entries outside the main diagonal are all zero. In other words, an $n \times n$ matrix $\Delta = (\delta_{i,j})$ is diagonal if $\delta_{i,j}=0$, $i \neq j, 1 \leq i,j \leq n$.
For example, $R_z (\theta)$ gate is a diagonal $2 \times 2$ matrix. A generic diagonal gate over $n$ qubits is represented by a $2^n \times 2^n$ unitary diagonal matrix. A diagonal gate over $n$ qubits is denoted as $\Delta_n$ throughout this paper.

A block matrix is a partition of a matrix into rectangular smaller matrices called blocks. A block diagonal matrix is a block square matrix whose main diagonal blocks are square matrices. Hence, the blocks outside the main diagonal are zero matrices. A block diagonal matrix has the form shown in (\ref{Eq:blockdig}) where each $U_i(2^m)$ is a $2^m\times 2^m$ matrix.
The inverse of a block diagonal matrix is also block diagonal which is composed of the inverse of each block.
A quantum gate $B^s_{\tau}(U(2^m))$ over $n$ qubits with $m$ targets denoted by the set $\tau$ and $s=n-m$ select qubits on the most significant qubits has a block diagonal matrix \cite{Shende06} represented by $B(U(2^m))$ in this paper. In the case of only one target line (i.e., $m=1$), the specific target line is shown instead of the set $\tau$. From the quantum circuit point of view, for a gate $B^s_{\tau}(U(2^m))$ over $n$ qubits with matrix $B(U(2^m))$ shown in (\ref{Eq:blockdig}), the matrix of $U_i(2^m)$ ($1 \leq i \leq 2^{n-m}$) is applied to the target qubits where the index $i$ depends on the values of $s=n-m$ select qubits \cite{Shende06}. In this paper, a select qubit in a circuit is denoted by $\Box$ as in \cite{Shende06}.
\begin{equation}\label{Eq:blockdig}
B(U(2^m)) = \left( {\begin{array}{*{20}c}
   {U_1(2^m) } & 0 &  \cdots  & 0  \\
   0 & {U_2(2^m) } &  \cdots  & 0  \\
    \vdots  &  \vdots  &  \ddots  &  \vdots   \\
   0 & 0 &  \cdots  & {U_{2^{n-m}}(2^m)}  \\
\end{array}} \right)
\end{equation}

The unitary matrix implemented by several gates acting on different qubits independently can be calculated by the tensor product $\otimes$ of their matrices. Two or more quantum gates can be cascaded to construct a quantum circuit. For a set of $k$ gates $g_1$, $g_2$, ..., $g_k$ cascaded in a quantum circuit $C$ in sequence, the matrix of $C$ can be calculated as $M_k M_{k-1} ... M_1$ where $M_i$ is the matrix of the $i^{th}$ gate ($1\leq i \leq k$).

Quantum circuits are often synthesized using a ``basic gate" library \cite{Barenco95} which contains CNOT and one-qubit gates. In contrast, an ``elementary gate" library was used in \cite{bullock-2004-4} which contains CNOT and one-qubit rotation gates. The gate CNOT acts on two qubits (control and target) where the state of the target qubit is inverted if the control qubit holds the state $|1\rangle$. A CNOT gate with control $c$ and target $t$ is denoted as $C^{c,t}$ in this paper. Since every gate in the elementary library is a basic gate and every basic gate can be decomposed into at most three elementary gates (see Section \ref{subsec:identities}),
the $\Omega$(gate count), O(gate count), and $\Theta$(gate count) in either gate library are identical. In this paper, the basic gate library is applied to synthesize a given unitary matrix as discussed in Section \ref{sec:our_method}.

\subsection{Quantum Circuit Decomposition} \label{subsec:identities}

Two quantum circuits are equivalent if matrix products of their gates are identical. In order to synthesize a given unitary matrix, equivalent circuits may be applied to simplify the circuit. To do this, various quantum circuit identities have been proposed in recent years (for examples see \cite{tucci-2004}, \cite{Nielsen00}, \cite{lomont-2003}, \cite{Shende06}).

An arbitrary one-qubit gate $U(2)$ can be decomposed into $R_z$ and $R_y$ rotation gates (called ZYZ decomposition) as shown in (\ref{Eq:ZYZ}) \cite{Barenco95}. Hence, a one-qubit computation from the basic gate library can be implemented as a
sequence of at most three gates from the elementary gate library.
\begin{equation}\label{Eq:ZYZ}
U(2)  = R_z (\alpha )R_y (\beta )R_z (\lambda )
\end{equation}

Cosine-sine decomposition \cite{Paige} for a $2^n \times 2^n$ unitary matrix $U$ can be expressed by (\ref{Eq:CSmath}) where $L_1$, $L_2$, $R_1$, and $R_2$ are unitary $2^{n-1} \times 2^{n-1}$ matrices and $C$ and $S$ are unitary $2^{n-1} \times 2^{n-1}$ diagonal matrices with real elements such that $C^2 + S^2 = I_{n-1}$ ($I_{n-1}$ is the identity matrix over $n-1$ qubits).
\begin{equation}\label{Eq:CSmath}
U =  {\left( {\begin{array}{*{20}c}
   {L_1 } & 0  \\
   0 & {L_2 }  \\
\end{array}} \right)} {\left( {\begin{array}{*{20}c}
   C & S  \\
   { - S} & C  \\
\end{array}} \right)}{\left( {\begin{array}{*{20}c}
   {R_1 } & 0  \\
   0 & {R_2 }  \\
\end{array}} \right)}
\end{equation}

In the following, two quantum synthesis algorithms which are based on the cosine-sine decomposition method are described and used in the rest of the paper.
\subsubsection{CSD} \label{subsec:csd}
In \cite{Mottonen-2004}, a decomposition algorithm, called CSD, was proposed which uses the cosine-sine decomposition to decompose a generic $U(2^n)$. The CSD decomposition was further improved in \cite{Bergholm:2004} where the cosine-sine decomposition is recursively applied on each block diagonal gate. This process is stopped when there are only one-qubit block diagonal gates. More precisely, the CSD decomposition of a generic $U(2^n)$ matrix can be expressed by (\ref{Eq:CSD}) where $\gamma(i)$ is the ruler function introduced in \cite{guy} ($\gamma(i)$+1 is the position of the least significant nonzero bit in the binary representation of $i$).
\begin{equation}\label{Eq:CSD}
U(2^n)= B_{n}^{n-1} (U(2))\prod\limits_{i=1}^{2^{n-1}-1} {B_{n-\gamma(i)}^{n-1} (R_y)B_{n}^{n-1} (U(2))}
\end{equation}

Each block diagonal gate that appears in (\ref{Eq:CSD}) can also be decomposed into a circuit which contains three block diagonal gates of smaller sizes and one CNOT gate \cite{Bergholm:2004}. Equation (\ref{Eq:BD}) shows the decomposition for a $B_{n}^{n-1} (U(2))$.
\begin{equation}\label{Eq:BD}
B_{n}^{n-1}(U(2)) = B_{n}^{n-2} (U(2))C^{1,n} B_{n}^{n-2}(U(2))B_{1}^{n-1}(R_z)
\end{equation}

\begin{figure}[h!]
	\centering
        \input{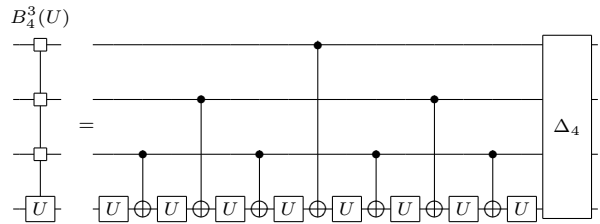}
		\caption{Decomposition of a block diagonal gate of size 4 \cite{Bergholm:2004}. Each $U$ is a generic $U(2)$ gate.}
	\label{Fig:CSD4}
\end{figure}

Applying the recursive decompositions given in (\ref{Eq:BD}) leads to a circuit with $2^{n-1}-1$ CNOTs and $2^{n-1}$ one-qubit gates followed by a diagonal gate $\Delta_n$ where $\Delta_n$ can also be decomposed into $2^n-2$ CNOTs \cite{bullock-2004-4}. Figure \ref{Fig:CSD4} illustrates the result of \cite{Bergholm:2004} for a block diagonal circuit with 4 qubits.

In addition, the authors of \cite{Bergholm:2004} used the notation given in (\ref{Eq:T-3}) and converted (\ref{Eq:CSD}) to (\ref{Eq:CSD2}) by replacing (\ref{Eq:T-3}) in (\ref{Eq:CSD}) and merging each $\Delta_n$ with its next block diagonal gate. Note that the notation $\sim$ over a block diagonal gate means that this gate can be implemented up to a diagonal gate. Doing this could reduce the number of gates resulted from the decomposition of a unitary gate.
\begin{equation}\label{Eq:T-3}
B_n^{n - 1} (U(2)) = \Delta _n \tilde B_n^{n - 1} (U(2))
\end{equation}
\begin{equation}\label{Eq:CSD2}
U(2^n)= \Delta_n \tilde B_{n}^{n-1} (U(2))\prod\limits_{i=1}^{2^{n-1}-1} {\tilde B_{n-\gamma(i)}^{n-1} (U(2))\tilde B_{n}^{n-1} (U(2))}
\end{equation}

\subsubsection{QSD}
Since the left and the right matrices in (\ref{Eq:CSmath}) are block diagonals, they can be represented as block diagonal gate $B^1_\tau(U(2^{n-1}))$. In addition, the middle matrix in (\ref{Eq:CSmath}) can be represented by $B^1_{n-1}(R_y)$. Therefore, (\ref{Eq:CSmath}) can be rewritten as (\ref{Eq:QSD-1}).
\begin{equation}\label{Eq:QSD-1}
U(2^n)= B_{\tau}^1(U(2^{n-1})) B_{1}^{n-1}(R_y) B_{\tau}^1(U(2^{n-1}))
\end{equation}

The authors of \cite{Shende06} showed that each block diagonal gate $B_{\tau}^1(U(2^{n-1}))$ in (\ref{Eq:QSD-1}) can be decomposed into two generic unitary matrices and a specific block diagonal matrix as illustrated in (\ref{Eq:DeMuxMath}) and (\ref{Eq:QSD-2}).
\begin{equation}\label{Eq:DeMuxMath}
B(U(2^{n-1}))
= \left( {\begin{array}{*{20}c}
   {U_1 } & 0  \\
   0 & {U_1 }  \\
\end{array}} \right){\left( {\begin{array}{*{20}c}
   D & 0  \\
   0 & D^\dagger \\
\end{array}} \right)}\left( {\begin{array}{*{20}c}
   {U_2 } & 0  \\
   0 & {U_2 }  \\
\end{array}} \right)
\end{equation}

\begin{equation}\label{Eq:QSD-2}
B_{\tau}^1(U(2^{n-1})) = U(2^{n-1}) B_1^{n-1}(R_z) U(2^{n-1})
\end{equation}

By using (\ref{Eq:QSD-1}) and (\ref{Eq:QSD-2}), the method of \cite{Shende06} applies the quantum Shannon decomposition (QSD) shown in (\ref{Eq:QSD-3}) to synthesize a given unitary matrix. The recursion is continued till $U(4)$. Next, an optimal decomposition of $U(4)$ with three CNOT gates \cite{Shende-1-2004} is used.
\begin{equation}\label{Eq:QSD-3}
\begin{array}{l}
 U(2^n) = U(2^{n - 1})B_1^{n - 1} (R_z )U(2^{n - 1})B_1^{n - 1} (R_y ) U(2^{n - 1})B_1^{n - 1} (R_z )U(2^{n - 1}) \\
 \end{array}
\end{equation}

\begin{figure}[h!]
	\centering
        \input{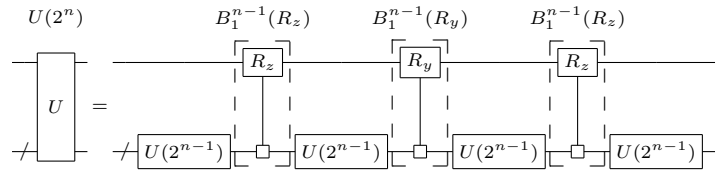}
		\caption{Applying QSD \cite{Shende-1-2004} on a $U(2^n)$ matrix }
	\label{Fig:QSD}
\end{figure}

The complete circuit diagram of (\ref{Eq:QSD-3}) is shown in Fig. \ref{Fig:QSD} where $B_1^{n-1} (R_z)$ and $B_1^{n-1} (R_y)$ gates are illustrated by dashed lines. In this paper, we use slash to denote that a given line may carry an arbitrary number of qubits. In addition, for a circuit with $n$ qubits, circuit lines are numbered from $1$ (top) to $n$ (bottom).

\section{The Proposed Synthesis Method} \label{sec:our_method}
In this section, a synthesis approach for quantum circuits is introduced.
The proposed approach is based on the CSD and QSD methods.

\subsection{The Basic Block} \label{sec:BB}
\begin{figure}[h!]
	\centering
        \input{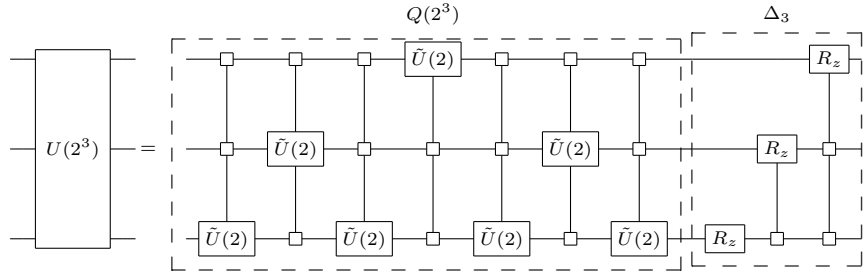}
		\caption{The proposed basic block for 3 qubits.}
	\label{Fig:Basic}
\end{figure}

In this part, a basic block is introduced which is used in the proposed decomposition method. Consider a generic unitary matrix over $l$ qubits and the CSD method is used to synthesize it as discussed in Section \ref{subsec:csd}. The resulted circuit without the last $\Delta_l$ gate is considered as the basic block in the proposed decomposition method. The proposed basic block is shown in (\ref{Eq:Basic-1}) denoted as $Q(2^l)$. The complete equation for $Q(2^l)$ is given in (\ref{Eq:Basic-2}). The CSD decomposition for $U(2^3)$ and our basic block are illustrated in Fig. \ref{Fig:Basic}. {The rightmost sequence of block diagonal $R_z$ gates corresponds to a single $\Delta_3$ gate.}

\begin{equation}\label{Eq:Basic-1}
U(2^l ) = \Delta _l Q(2^l)\\
\end{equation}

\begin{equation}\label{Eq:Basic-2}
Q(2^l)= \tilde B_{l}^{l-1} (U(2))\prod\limits_{i=1}^{2^{l-1}-1} {\tilde B_{l-\gamma(i)}^{l-1} (U(2))\tilde B_{l}^{l-1} (U(2))}
\end{equation}

\subsection{Block-Based Quantum  Decomposition (BQD)} \label{sec:synmethod}
\begin{figure}[h!]
	\centering
        \input{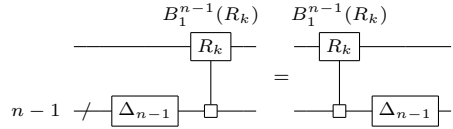}
		\caption{Application of Proposition \ref{Lemma:1} on $n$ qubits.}
	\label{Fig:Lemma1}
\end{figure}

In this subsection, the proposed decomposition method is introduced. The following proposition is used in the rest of this section to remove unnecessary gates.
\vspace*{12pt}
\noindent
\begin {prop} \label{Lemma:1}
$(\Delta_l \otimes I_{n-l}) B_{1}^{n-1} (R_k) = B_{1}^{n-1} (R_k) (\Delta_l \otimes I_{n-l}) $ where $l\leq n-1$ and $k$ = $z$,$y$.
\end {prop}
\vspace*{12pt}
\noindent
As there is no overlap between the target line of $B_{1}^{n-1} (R_k)$ and $\Delta_l$ for $l\leq n-1$ and $k$ = $z$,$y$, Proposition \ref{Lemma:1} is held.
Figure \ref{Fig:Lemma1} illustrates the result of Proposition \ref{Lemma:1} on $n$ qubits for $\Delta_{n-1}$.
\vspace*{12pt}
\noindent

\vspace*{12pt}
\noindent

Reconsider the QSD method shown in (\ref{Eq:QSD-3}). It can be verified that the recursive application of (\ref{Eq:QSD-3}) up to $l$ qubits leads to (\ref{Eq:QSD-4}) where each $U(2^l)$ is denoted as $U$ to save space.
\begin{equation}\label{Eq:QSD-4}
\begin{array}{l}
 U(2^n ) = UB_{n - l}^l (R_z )UB_{n - l}^l (R_y )UB_{n - l}^l (R_z )U B_{n - l-1}^{l+1} (R_z )...B_{n - l-1}^{l+1} (R_y )...B_{n - l-1}^{l+1} (R_z )... \\
 \,\,\,\,\,\,\,\,\,\,\,\,\,\,\,\,\,\,\,\,\,B_1^{n - 1} (R_z )...B_1^{n - 1} (R_y )...B_1^{n - 1} (R_z )... \\
 \,\,\,\,\,\,\,\,\,\,\,\,\,\,\,\,\,\,\,\,\,B_{n - l-1}^{l+1} (R_z )...B_{n - l-1}^{l+1} (R_y )...B_{n - l-1}^{l+1} (R_z )UB_{n - l}^l (R_z )UB_{n - l}^l (R_y )UB_{n - l}^l (R_z )U \\
 \\
 \end{array}
\end{equation}

Based on (\ref{Eq:Basic-1}), each $U(2^l)$ in (\ref{Eq:QSD-4}) can be replaced by $\Delta _l Q(2^l)$. The result is given in (\ref{Eq:Decomp-1}) where each $Q(2^l)$ is denoted as $Q$.
\begin{equation}\label{Eq:Decomp-1}
\begin{array}{l}
 U(2^n ) = \\
 \,\,\,\,\,\,\,\,\,\Delta_l Q B_{n - l}^l (R_z ) \Delta_l Q B_{n - l}^l (R_y ) \Delta_l Q B_{n - l}^l (R_z ) \Delta_l Q B_{n - l-1}^{l+1} (R_z )...B_{n - l-1}^{l+1} (R_y )...B_{n - l-1}^{l+1} (R_z )... \\
 \,\,\,\,\,\,\,\,\,B_1^{n - 1} (R_z )...B_1^{n - 1} (R_y )...B_1^{n - 1} (R_z )... \\
 \,\,\,\,\,\,\,\,\,B_{n - l-1}^{l+1}(R_z )...B_{n - l-1}^{l+1} (R_y )...B_{n - l-1}^{l+1} (R_z )\Delta_l Q B_{n - l}^l (R_z ) \Delta_l Q B_{n - l}^l (R_y ) \Delta_l Q B_{n - l}^l (R_z ) \Delta_l Q \\
 \\
 \end{array}
\end{equation}

{Now, direct applying of Proposition \ref{Lemma:1} on (\ref{Eq:Decomp-1}) moves all $\Delta_l$ gates to the right side of block diagonal gates. Then, each $\Delta_l$ gate can be merged with its adjacent $Q(2^l)$ gate which leads to a $U(2^l)$ gate. The resulted $U(2^l)$ gate can also be replaced by $\Delta _l Q(2^l)$. Continuing this process leads to (\ref{Eq:Decomp-2}).}

\begin{equation}\label{Eq:Decomp-2}
\begin{array}{l}
 U(2^n ) = \\
 \,\,\,\,\,\,\,\,\,\Delta_l Q B_{n - l}^l (R_z )Q B_{n - l}^l (R_y )Q B_{n - l}^l (R_z )Q B_{n - l-1}^{l+1} (R_z )...B_{n - l-1}^{l+1} (R_y )...B_{n - l-1}^{l+1} (R_z )... \\
 \,\,\,\,\,\,\,\,\,B_1^{n - 1} (R_z )...B_1^{n - 1} (R_y )...B_1^{n - 1} (R_z )... \\
 \,\,\,\,\,\,\,\,\,B_{n - l-1}^{l+1}(R_z )...B_{n - l-1}^{l+1} (R_y )...B_{n - l-1}^{l+1} (R_z )Q B_{n - l}^l (R_z )Q B_{n - l}^l (R_y )Q B_{n - l}^l (R_z )Q \\
 \\
 \end{array}
\end{equation}

\begin{figure}[h!]
	\centering
        \input{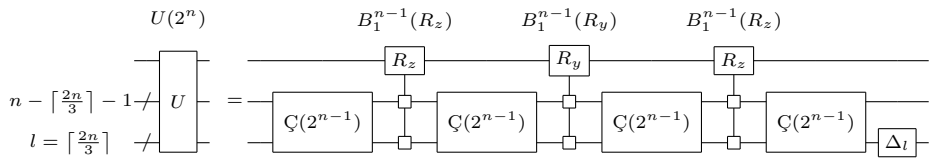}
		\caption{The proposed decomposition method}
	\label{Fig:LQD}
\end{figure}

{
Based on the above equations, the proposed block-based quantum decomposition (BQD) method is given in (\ref{Eq:Decomp-5}) where each $\c{C}(2^i)$ is denoted as (\ref{Eq:Decomp-4}). The circuit diagrams of (\ref{Eq:Decomp-5}) and (\ref{Eq:Decomp-4}) are illustrated in Fig. \ref{Fig:LQD} and Fig. \ref{Fig:Q_i}, respectively. Figure \ref{Fig:Decomp-6} shows the result of BQD method for $U(2^6)$ where $l=4$.
}
{
\begin{equation}\label{Eq:Decomp-5}
U(2^n ) = \Delta _l \c{C}(2^{n - 1} )B_1^{^{n - 1} } (R_z )\c{C}(2^{n - 1} )B_1^{^{n - 1} } (R_y)\c{C}(2^{n - 1} )B_1^{^{n - 1} } (R_z )\c{C}(2^{n - 1} )
\end{equation}
}
{
\begin{equation}\label{Eq:Decomp-4}
\c{C}(2^i ) = \left\{ \begin{array}{l}
\c{C}(2^{i - 1} )B_1^{^{i - 1} } (R_z )\c{C}(2^{i - 1} )B_1^{^{i - 1} } (R_y)\c{C}(2^{i - 1} )B_1^{^{i - 1} } (R_z )\c{C}(2^{i - 1} )\,\,\,\,\, l < i \leq n\\
Q(2^l)\,\,\,\,\,\,\,\,\,\,\,\,\,\,\,\,\,\,\,\,\,\,\,\,\,\,\,\,\,\,\,\,\,\,\,\,\,\,\,\,\,\,\,\,\,\,\,\,\,\,\,\,\,\,\,\,\,\,\,\,\,\,\,\,\,\,\,\,\,\,\,\,\,\,\,\,\,\,\,\,\,\,\,\,\,\,\,\,\,\,\,\,\,\,\,\,\,\,\,\,\,\,\,\,\,\,\,\,\,\,\,\,\,\,\,\,\,\,\,\,\,\,\,\,\,\,\,\,\,\,\,\,\,\,\,\,\,\,\,\,\,\,\,\,i=l\\
\end{array} \right.
\end{equation}
}
\begin{figure}[h!]
	\centering
        \input{circuits/Q_i.tex}
		\caption{The circuit diagram of (\ref{Eq:Decomp-4})}
	\label{Fig:Q_i}
\end{figure}
{
Note that only one $\Delta_l$ exists after final decomposition stopped at level $l$. To determine $l$, i.e., \emph{decomposition level}, number of CNOT gates and one-qubit gates should be evaluated as done in the following subsections.
}
\begin{figure}[h!]
	\centering
        \input{circuits/Decomp-6.tex}
		\caption{Decomposition of a $U(2^6)$ using BQD method. $Q(2^4)$ is denoted as $Q$.}
	\label{Fig:Decomp-6}
\end{figure}
\subsection{Synthesis Cost}\label {syn_cost}
\subsubsection{CNOT cost}\label {syn_cost_cnot}

Equation (\ref{Eq:cnot-0}) shows the recursion formula for the number of CNOT gates for the circuit of (\ref{Eq:Decomp-4}) where $2^{i - 1}$ is the number of CNOT gates for each $R_z$ gate \cite{Shende06}. {In addition, the block diagonal $R_y$ gate in Fig. \ref{Fig:Q_i} can be transformed in the decomposition into a block diagonal $U(2)$ rotation gate up to a diagonal gate. As a result, it can be implemented by one less CNOT compared with the block diagonal $R_z$ gate \cite{Mottonen05}, \cite{Shende06}.}

\begin{equation}\label{Eq:cnot-0}
C_{\c{C}_{i}}  = \left\{ \begin{array}{l}
4C_{\c{C}_{i - 1} }  + 3 \times 2^{i - 1} - 1 \,\,\,\,\,\,\,\,\, l < i \leq n\\
C_{Q_l}\,\,\,\,\,\,\,\,\,\,\,\,\,\,\,\,\,\,\,\,\,\,\,\,\,\,\,\,\,\,\,\,\,\,\,\,\,\,\,\,\,\,\,\,\,\,\,\,\,\,\,\,\,\,\,\,\,\,\,\, i=l\\
\end{array} \right.
\end{equation}

{Applying the recursion formula of (\ref{Eq:cnot-0}) leads to (\ref{Eq:cnot-1}).}

\begin{equation}\label{Eq:cnot-1}
C_{\c{C}_{i}}  = 4^{i - l} (C_{Q_l}  + 3 \times 2^{l - 1} ) - 3 \times 2^{i - 1} - \frac{{4^{n - l}  - 1}}{3}\,\,\,\,\, l < i \leq n
\end{equation}

Now consider the proposed BQD method shown in (\ref{Eq:Decomp-5}). Equation (\ref{Eq:cnot-2.0}) shows the number of CNOT gates in the proposed method where $2^l  - 2$ is the number of CNOT gates for $\Delta_l$ \cite{bullock-2004-4}.

\begin{equation}\label{Eq:cnot-2.0}
C_{U_{n}}  = C_{\c{C}_{n}} + 2^l  - 2
\end{equation}

{One more CNOT gate can be eliminated in $Q(2^l)$ $\Delta_l$ \cite{Bergholm:2004}. This comes from the fact that $Q(2^l)$ $\Delta_l$ can be written as a sequence of $\tilde B_{i}^{l-1}(U(2))$ gates, where $i\in\{l, l+1, ..., n-1,n\}$, followed by a chain of block diagonal $R_z$ gates (see Fig. \ref{Fig:Basic}). The single rotation gate in the chain can be absorbed into a $U(2)$ gate appeared after the decomposition of $\tilde B_{n}^{l-1} (U(2))$ gate. Then, the sub-circuit $I \otimes U(2) C^{1,2} I \otimes U(2)$, resulted from the decomposition of $\tilde B_{n}^{l-1} (U(2))$ gate, followed by a $B_{n-1}^{1}(R_z)$ gate constructs a $B_{n}^{1} (U(2))$ gate, which can be implemented by two CNOT gates \cite{Shende06}, one less than the number of CNOT gates required for the implementation of $I \otimes U(2) C^{1,2} I \otimes U(2) B_{n-1}^{1}(R_z)$ sub-circuit. Equation (\ref{Eq:cnot-2}) shows the final result.}

\begin{equation}\label{Eq:cnot-2}
\begin{array}{l}
C_{U_{n} }  = 4^{n - l} (C_{Q_l}  + 3 \times 2^{l - 1} ) - 3 \times 2^{n - 1} - \frac{{4^{n - l} - 1}}{3} + 2^l  - 3\\
\end{array}
\end{equation}

The number of CNOT gates for $Q(2^l)$ (see (\ref{Eq:Basic-2})) can be computed by (\ref{Eq:cnot-3}) where $2^l  - 1$ is the number of block diagonal gates in $Q(2^l)$ each of which can be implemented by $2^{l - 1}  - 1$ CNOT gates \cite{Bergholm:2004}.
\begin{equation}\label{Eq:cnot-3}
C_{Q_l}  = (2^l  - 1) \times (2^{l - 1}  - 1)
\end{equation}

Altogether, the number of CNOT gates for the proposed BQD method is shown in (\ref{Eq:cnot-4}) for the decomposition level $l$.

\begin{equation}\label{Eq:cnot-4}
C_{U_{n} } = 4^n (\frac{2}{{3 \times 4^l }} + \frac{1}{2}) - 3 \times 2^{n - 1}  + 2^l  - \frac{8}{3}
\end{equation}

\subsubsection{One-qubit cost}
To count the number of one-qubit gates in the proposed decomposition method, an analysis similar to the one performed for CNOT cost can be used. Equation (\ref{Eq:onequb-1}) shows the number of one-qubit gates for the proposed BQD method where $2 \times \frac{{4^{n - l}  - 1}}{3}$ is the number of one-qubit gates merged between ${B}(R_y)$ and ${B}(R_z)$ gates at each recursion step. $2^l  - 1$ is the number of one-qubit gates needed for diagonal gate $\Delta_l$. Besides, $l$ more gates can be eliminated at the last $U(2^l)$ gate decomposed using CSD method \cite{Mottonen05}.
\begin{equation}\label{Eq:onequb-1}
O_{U_{n} }  = 4^{n - l} (O_{Q _l }  + 3 \times 2^{l - 1} ) - 3 \times 2^{n - 1}  - (2 \times \frac{{4^{n - l}  - 1}}{3}) + (2^l  - 1) - l
\end{equation}

Number of one-qubit gates for $Q(2^l)$ is computed by (\ref{Eq:onequb-2}).
\begin{equation}\label{Eq:onequb-2}
O_{Q_l}  = (2^l  - 1) \times (2^{l - 1}  - 1)
\end{equation}

Replacing (\ref{Eq:onequb-2}) in (\ref{Eq:onequb-1}) and doing some simplifications lead to (\ref{Eq:onequb-3}) for decomposition level $l$.
\begin{equation}\label{Eq:onequb-3}
O_{U_{n} } = 4^n ( \frac{-2}{{3 \times 4^l }} + \frac{1}{{2^l }} + \frac{1}{2}) - 3 \times 2^{n - 1}  + 2^l  - l - \frac{1}{3}
\end{equation}

\section{Result Comparison}\label{sec:results}
In this section, the results of the proposed decomposition approach is presented. To this end, the effect of the decomposition level on the synthesis cost in the BQD method is discussed first. Table \ref{table:level-1} and Table \ref{table:level-2} show the number of CNOT gates and one-qubit gates for $4\le n$ and $3 \le l \le n$ up to $n=12$, respectively. Additionally, the total number of gates in BQD method for each decomposition level is shown in Table \ref{table:level-3}. Note that the synthesis results for $l=n$ in the proposed decomposition method (gray cells) are the same as the ones produced by the improved CSD method (i.e., \cite{Bergholm:2004}). The results of the QSD method \cite{Shende06} are also shown in these tables.

\renewcommand{\arraystretch}{1.3}

\begin{table}[h]
\caption{Number of CNOT gates in the proposed BQD, improved CSD \cite{Bergholm:2004} (gray cells), and QSD \cite{Shende06} methods for $4\le n \le 12$.}
\label{table:level-1}
\centering
\scriptsize
\begin{tabular}{|c|c|c|c|c|c|c|c|c|c|}
\hline
Decomposition& \multicolumn{9}{c|}{Circuit lines ($n$)}  \\
Level ($l$)   & 4 & 5 & 6 & 7 & 8 & 9 & 10 & 11 & 12   \\
\hline
\hline
 3 & \textbf{112} & \textbf{480} & 2000 & 8176 & 33072 & 133040 & 533680 & 2137776 &8557232 \\
\hline
{4} & \multicolumn{0}{>{\columncolor[gray]{0.5}}l|}{118} & {\textbf{480}} &{\textbf{1976}}	& {8056}	 & {32568} & {131000} & {525496}	&{2105016} &{8426168}\\
\hline
5 & & \multicolumn{0}{>{\columncolor[gray]{0.5}}l|}{494} & 1984	&\textbf{8040}	 &\textbf{32456}&	130504&	523464 &	 2096840 &8393416 \\
\hline
6 & & & \multicolumn{0}{>{\columncolor[gray]{0.5}}l|}{2014} & 8064	&\textbf{32456}	 &\textbf{130408}&	522984&	2094824 &8385256\\
\hline
7 & & & & \multicolumn{0}{>{\columncolor[gray]{0.5}}l|}{8126} & 32512 & 130440	 &\textbf{522920}	&\textbf{2094376} & 8383272 \\
\hline
8 & & & & & \multicolumn{0}{>{\columncolor[gray]{0.5}}l|}{32638} & 130560	&523016	 &\textbf{2094376} &  \textbf{8382888}\\
\hline
9 & & & & & & \multicolumn{0}{>{\columncolor[gray]{0.5}}l|}{130814} &523264	&2094600 & 8383016\\
\hline
10 & & & & & & & \multicolumn{0}{>{\columncolor[gray]{0.5}}l|}{523774} & 2095104 &8383496\\
\hline
11 & & & & & & & & \multicolumn{0}{>{\columncolor[gray]{0.5}}l|}{2096126} & 8384512\\
\hline
12 & & & & & & & & & \multicolumn{0}{>{\columncolor[gray]{0.5}}l|}{8386558}  \\
\hline
\hline
\hline
\cite{Mottonen05}, QSD \cite{Shende06} & 100 & 444 & 	1868	& 7660 & 31020 & 124844 & 500908  & 2006700 & 8032940 \\
\hline
\end{tabular}

\end{table} 
\renewcommand{\arraystretch}{1.3}

\begin{table}[h]
\caption{Number of one-qubit gates in the proposed BQD, improved CSD \cite{Bergholm:2004} (gray cells), and QSD \cite{Shende06} methods for $4\le n \le 12$.}
\label{table:level-2}
\centering
\scriptsize
\begin{tabular}{|c|c|c|c|c|c|c|c|c|c|}
\hline
Decomposition& \multicolumn{9}{c|}{Circuit lines ($n$)}  \\
Level ($l$)   & 4 & 5 & 6 & 7 & 8 & 9 & 10 & 11 & 12  \\
\hline
\hline
 3 & \textbf{138}&	586&	2426&	9882&	39898&	160346&	642906&	2574682 & 10304858\\
\hline
 4 & \multicolumn{0}{>{\columncolor[gray]{0.5}}l|}{131} & \textbf{537}	&\textbf{2209}	&8993	&36321	&146017	&585569	 &2345313 &9387361\\
\hline
5 & & \multicolumn{0}{>{\columncolor[gray]{0.5}}l|}{522} & 2104&	\textbf{8528} &	34416&	138352&	554864&	2222448 &8895856\\
\hline
6 & & & \multicolumn{0}{>{\columncolor[gray]{0.5}}l|}{2073} & 8311&	\textbf{33455}&	\textbf{134415} &	539023&	2158991 &8641935\\
\hline
7 & & & & \multicolumn{0}{>{\columncolor[gray]{0.5}}l|}{8248} & 33014&	132462&	\textbf{531022}	&2126798 & 8512974\\
\hline
8 & & & & & \multicolumn{0}{>{\columncolor[gray]{0.5}}l|}{32887} & 131573	&527085&	\textbf{2110669} &   \textbf{8448077}\\
\hline
9 & & & & & & \multicolumn{0}{>{\columncolor[gray]{0.5}}l|}{131318} &525300	&2102764 &  8415692\\
\hline
10 & & & & & & & \multicolumn{0}{>{\columncolor[gray]{0.5}}l|}{524789} & 2099187 &8399851 \\
\hline
11 & & & & & & & & \multicolumn{0}{>{\columncolor[gray]{0.5}}l|}{2098164} & 8392690\\
\hline
12 & & & & & & & & & \multicolumn{0}{>{\columncolor[gray]{0.5}}l|}{8390643} \\
\hline
\hline
\hline
QSD \cite{Shende06}  & 157	&677	&2805	&11413	&46037	&184917	&741205 & 2967893 & 11877717 \\
\hline
\end{tabular}
\end{table}

\renewcommand{\arraystretch}{1.3}

\begin{table}[h]
\caption{Total number of gates in the proposed BQD, improved CSD \cite{Bergholm:2004} (gray cells), and QSD \cite{Shende06} methods for $4\le n \le 12$.}
\label{table:level-3}
\centering
\scriptsize
\begin{tabular}{|c|c|c|c|c|c|c|c|c|c|}
\hline
Decomposition& \multicolumn{9}{c|}{Circuit lines ($n$)}  \\
Level ($l$)   & 4 & 5 & 6 & 7 & 8 & 9 & 10 & 11 & 12   \\
\hline
\hline
 3 & \textbf{250} & 1066	&4426	&18058	&72970	&293386	&1176586	&4712458	&18862090\\
\hline
{4} & \multicolumn{0}{>{\columncolor[gray]{0.5}}l|}{249}	&\textbf{1017}	&\textbf{4185}	&17049	&68889	&277017	 &1111065	&4450329	 &17813529\\
\hline
5 & & \multicolumn{0}{>{\columncolor[gray]{0.5}}l|}{1016}	&4088	&\textbf{16568}	&66872	&268856	&1078328	&4319288	 &17289272\\
\hline
6 & & & \multicolumn{0}{>{\columncolor[gray]{0.5}}l|}{4087}	&16375	&\textbf{65911}	&\textbf{264823}	&1062007	 &4253815	&17027191\\
\hline
7 & & & & \multicolumn{0}{>{\columncolor[gray]{0.5}}l|}{16374}	&65526	&262902	&\textbf{1053942}	&4221174	 &16896246\\
\hline
8 & & & & & \multicolumn{0}{>{\columncolor[gray]{0.5}}l|}{65525}	&262133	&1050101	&\textbf{4205045}	 &\textbf{16830965}\\
\hline
9 & & & & & & \multicolumn{0}{>{\columncolor[gray]{0.5}}l|}{262132}	&1048564	&4197364	&16798708\\
\hline
10 & & & & & & &\multicolumn{0}{>{\columncolor[gray]{0.5}}l|}{1048563}	&4194291	&16783347\\
\hline
11 & & & & & & & & \multicolumn{0}{>{\columncolor[gray]{0.5}}l|}{4194290}	&16777202\\
\hline
12 & & & & & & & & & \multicolumn{0}{>{\columncolor[gray]{0.5}}l|}{16777201}\\
\hline
\hline
\hline
QSD \cite{Shende06}  & 257	&1121 	&4673 	&19073	&77057	&309761	&1242113 & 4974593 & 19910657 \\
\hline
\end{tabular}

\end{table} 

The results given in Table \ref{table:level-1}, Table \ref{table:level-2} and Table \ref{table:level-3} reveal the following facts regarding the behavior of the BQD method compared to the other approaches:

\begin{itemize}
  \item In terms of the number of CNOT gates, QSD and \cite{Mottonen05} work always better than BQD and BQD works better than CSD.
  \item In terms of the number of one-qubit gates, improved CSD is better than BQD and BQD is better than QSD and \cite{Mottonen05} all the time. Similar result is obtained in terms of total number of gates.
  \item As can be seen in Table \ref{table:level-3}, in $l=n-1$, the total number of gates produced by BQD is one gate more than the number of gates produced by improved CSD. However, in those cases, BQD produces fewer CNOT gates. Since CNOT cost is usually much more than the cost of one-qubit gates, BQD produces better circuits compared to the other methods as far as the total number of gates and technological limitations are concerned. The appropriate $l$ is $n-1$ for this case.
  \item While QSD and \cite{Mottonen05} lead to better CNOT cost compared to improved CSD and BQD, if one-qubit gate count is also concerned, BQD produces about 5\% more CNOT gates and about 28\% fewer one-qubit gates compared to QSD on average. Altogether, the total number of gates is improved by about 16\% in BQD compared to QSD on average. The best results in terms of the number of CNOT gates for each $n$ are boldfaced in all tables. In some cases, there are two different decompositions which have the same number of CNOT gates and different numbers of one-qubit gates. In those cases, decompositions with fewer one-qubit gates are selected. Percentages of improvement (BQD vs. QSD) for CNOT gates, one-qubit gates, and total gates are shown in Fig. \ref{fig:comp} for this case.

\begin{figure}[h]
	\centering
		\includegraphics[scale=0.8]{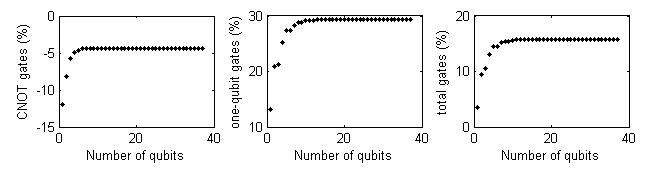}
		\caption{Percentages of improvement for CNOT, one-qubit and total gates (BQD vs. QSD)}
        \label {fig:comp}
\end{figure}

A comparison of the results produced by BQD and improved CSD reveals that for this case, BQD leads to fewer CNOT gates with the penalty of more one-qubit gates. Percentages of improvement (BQD vs. improved CSD) for CNOT gates and one-qubit gates are almost equal (with a negative sign). However, fewer CNOT gates is more desirable as discussed.

\item Producing fewer CNOT gates leads to better quantum circuits considering the current technological limitations. To find the exact gate counts in the proposed method, we set derivation of (\ref{Eq:cnot-4}), i.e., the number of CNOT gates for BQD method, with respect to $l$ to zero which leads to $l=\frac{2n+log_2(4/3)}{3}$. According to the results given in Table \ref{table:level-1} and Table \ref{table:level-2}, it can be verified that $l=\lceil 2n/3 \rceil$ leads to the achieved results.
          Therefore, the number of CNOT gates and one-qubit gates in the proposed BQD method can be calculated as (\ref{Eq:cnot-5}) and (\ref{Eq:onequb-4}), respectively.

    \begin{equation}\label{Eq:cnot-5}
    C_{U_{n}} =
    {\textstyle{1 \over 2}}4^n  - {\textstyle{3 \over 2}}2^n  + {\textstyle{2 \over 3}}4^{n - \left\lceil {2n/3} \right\rceil }  + 2^{\left\lceil {2n/3} \right\rceil }  - {\textstyle{8 \over 3}}
    \end{equation}

    \begin{equation}\label{Eq:onequb-4}
    O_{U_{n}} =
    {\textstyle{1 \over 2}}4^n  + 2^{2n - \left\lceil {2n/3} \right\rceil }  - {\textstyle{3 \over 2}}2^n  - {\textstyle{2 \over 3}}4^{n - \left\lceil {2n/3} \right\rceil }  + 2^{\left\lceil {2n/3} \right\rceil }  - \left\lceil {{\textstyle{{2n} \over 3}}} \right\rceil  - {\textstyle{1 \over 3}}
    \end{equation}

\end{itemize}

          According to the above discussion, we select $l=\lceil2n/3\rceil$ in the proposed decomposition method.
          Table \ref{table:comp1} and Table \ref{table:comp2} show the number of CNOT gates and one-qubit gates for different quantum synthesis methods.
Considering the fact that number of CNOT gates in the best reported synthesis methods \cite{Mottonen05,Shende06} is away from the lower bound by a factor of two, more improvement in the synthesis cost needs consideration of both CNOT and one-qubit gates. Simultaneous reduction of CNOT and one-qubit gates has also been recently pursued by other researchers \cite{Shende09}.

\renewcommand{\arraystretch}{1.3}

\begin{table}[t]
\caption{Comparison table for the number of CNOT gates in various quantum synthesis methods}
\label{table:comp1}
\centering
\scriptsize
\begin{tabular}{|l|c|c|c|c|c|c|c|c|c|l|}
\hline
Synthesis& \multicolumn{10}{c|}{Circuit lines}  \\
 method   & 2 & 3 & 4 & 5 & 6 & 7 & 8 & 9 & 10 & $n$ \\
\hline
\hline
\cite{Barenco95},\cite{George2001}& \multicolumn{9}{c|}{-} & $O(n^34^n)$ \\
\hline
\cite{knill95}& \multicolumn{9}{c|}{-} & $O(n4^n)$ \\
\hline

 QR \cite{vartiainen-2004} & \multicolumn{9}{c|}{-} & $O(4^n)$ \\
\hline
Original CSD \cite{Mottonen-2004} & 8	& 48 & 224 &	960 &	3968 &	16128 &	65024 &	261120 &	1046528 & $4^n - 2^{n+1}$ \\
\hline
Improved CSD \cite{Bergholm:2004}  & 4 & 26 & 118 & 494 & 2014 & 8126	& 32638 & 130814 & 523774 & ${\textstyle{1 \over 2}}4^n  - {\textstyle{1 \over 2}}2^n  - 2$ \\
\hline
\cite{Mottonen05}, QSD \cite{Shende06}  & 3 & 20 & 100 & 444 & 	1868	& 7660 & 31020 & 124844 & 500908 & $\frac{23}{48} 4^n  - \frac {3}{2} 2^n  + \frac{4}{3}$ \\
\hline
\textbf{BQD} & -	&-	&112	&480	&1976	&8040	&32456	&130408	&522920 & Eq. (\ref{Eq:cnot-5})\\
\hline
\hline

Lower Bound \cite{shende-2004}& 3 & 14 & 61 & 252 & 1020 & 4091 & 16378  &	65529& 262137 & $\left\lceil \frac{1}{4}(4^n  - 3n - 1) \right\rceil$ \\

\hline

\end{tabular}
\end{table}

\begin{table}[t]
\caption{Comparison table for the number of one-qubit gates}
\label{table:comp2}
\centering
\scriptsize
\begin{tabular}{|l|c|c|c|c|c|c|c|c|c|l|}
\hline
 Synthesis & \multicolumn{10}{c|}{Circuit lines}  \\
 method & 2 & 3 & 4 & 5 & 6 & 7 & 8 & 9 & 10 & $n$ \\

\hline
\hline

Improved CSD  \cite{Bergholm:2004}  & 7 & 32 & 131 & 522 & 2073 & 8248 & 32887 & 131318 & 524789 & $\frac{1}{2} 4^n  + \frac{1}{2} 2^n  - n - 1$ \\
\hline

QSD \cite{Shende06} & 7	&33	&157	&677	&2805	&11413	&46037	&184917	&741205 &${\textstyle{{{\rm 17}} \over {{\rm 24}}}}{\rm 4}^{ n} {\rm   -  }{\textstyle{{\rm 3} \over {\rm 2}}}{\rm 2}^{ n} {\rm   - }{\textstyle{{\rm 1} \over {\rm 3}}}$\\

\hline
\textbf{BQD} & -	&-	&138	&537	&2209	&8528	&33455	&134415 &531022 & Eq. (\ref{Eq:onequb-4})\\
\hline

\end{tabular}
\end{table}

\section{Nearest Neighbor Implementation} \label{sec:NN}
While several impressive physical realizations have been proposed for quantum computers, all of these technologies have serious intrinsic limitations which should be resolved in future \cite{Ross2008}. Among the different technological constraints, the limited interaction distance between qubits is one of the most common ones. Although arbitrary-distance interaction between qubits is possible in quantum computer technologies with moving qubits (for example in a photon-based system \cite{Photon}), some restrictions exist in other quantum technologies. Indeed, many physical quantum computer proposals only permit interactions between adjacent (nearest neighbor) qubits \cite{FDH:2004}. For example, liquid nuclear magnetic resonance (NMR) \cite{NMR}, and the original Kane model \cite{Kane} were designed based on the interactions between linear nearest neighbor (LNN) qubits.

In this section, the worst-case cost of BQD synthesis method for nearest neighbor architecture is studied. To this end, the structure of the proposed basic block, the block diagonal and the diagonal gates under LNN constraints should be evaluated. Reconsider the proposed basic block $Q(2^l)$ which has $2^l - 1$ block diagonal gates. Equation (\ref{Eq:nn0}) shows the number of adjacent CNOT gates for a block diagonal gate with $l-1$ control lines where $s$ is between $1$ and $\lceil\frac{l}{2}\rceil$ \cite{Bergholm:2004}.

\begin{equation}\label{Eq:nn0}
C_{B_t^{l - 1} }  = {\textstyle{5 \over 6}}2^l  + 2l - 6s \left\{ \begin{array}{l}
 {\frac{1}{3}}\,\,{\kern 1pt} \,\,l\,\,\rm{even} \\
 {\frac{5}{3}}{\kern 1pt} \,\,\,{\kern 1pt} {\kern 1pt} l\,\,\rm{odd} \\
 \end{array} \right.
\end{equation}

A comparison between the number of CNOT gates required to implement a block diagonal gate without the LNN constraint (i.e., $2^{l-1}-1$) and with this constraint (i.e., $O(\frac{5}{6} 2^l)$) reveals that the number of CNOT gates for a block diagonal gate is increased by a factor of $\frac{5}{3}$ in the LNN architectures.

Equation (\ref{Eq:nn1}) shows the number of adjacent CNOTs of a block diagonal gate with $n-1$ control lines where target is at the $1^{st}$ line (without the last $\Delta_n$ gate) in the LNN architectures \cite{Bergholm:2004}. The block diagonal $R_y$ gates produced in the QSD method can also be implemented by the same number of adjacent CNOTs as discussed in Subsection \ref{syn_cost}.

\begin{equation}\label{Eq:nn1}
C_{B_1^{n - 1} }  = {\textstyle{5 \over 6}}2^n  + 2n - \left\{ \begin{array}{l}
 {\frac{19}{3}}\,\,{\kern 1pt} \,\,n\,\,\rm{even} \\
 {\frac{23}{3}}{\kern 1pt} \,\,\,{\kern 1pt} {\kern 1pt} n\,\,\rm{odd} \\
 \end{array} \right.
\end{equation}
Equation (\ref{Eq:nn2}) shows the number of adjacent CNOTs required for block diagonal $R_z$ gates \cite{Bergholm:2004} which have been produced during QSD decomposition.

\begin{equation}\label{Eq:nn2}
C_{R_1^{n - 1}}  = {\textstyle{5 \over 6}}2^n  + 3n - \left\{ \begin{array}{l}
 {\frac{22}{3}}{\kern 1pt} \,\,\,\,n\,\,\rm{even} \\
 {\frac{23}{3}}{\kern 1pt} \,\,\,\,n\,\,\rm{odd} \\
 \end{array} \right.
\end{equation}
A comparison between the number of CNOT gates required to implement a block diagonal gate without the LNN constraint (i.e., $2^{n-1}-1$) and with this constraint (i.e., $O(\frac{5}{6} 2^n)$) reveals that the number of CNOTs for a block diagonal gate is increased by a factor of $\frac{5}{3}$ in the LNN architectures.

On the other hand, the last diagonal gate can be considered as a cascade of block diagonal rotations with $i \in \{1, \cdots, l-1\}$ controls and one target placed at the $1^{st}$ qubit \cite{bullock-2004-4}.
Therefore, the nearest neighbor implementation of $\Delta_l$ can be computed as (\ref{Eq:nn3}) where $C_{\Delta _l}$ shows the number of CNOT gates in $\Delta_l$ for LNN architectures. {Equation (\ref{Eq:nn3}) reveals that $C_{\Delta _l }$ is of $O(\frac{5}{3}2^l)$ for LNN architectures.}

\begin{equation}\label{Eq:nn3}
C_{\Delta _l }  = {\textstyle{5 \over 3}}2^l  + {\textstyle{3 \over 2}}l^2  - \left\{ \begin{array}{l}
 {\textstyle{{35} \over 6}}l - {\textstyle{{41} \over 3}}  \,\,\,\,n\,\,\rm{even}\\
 {\textstyle{{37} \over 6}}l - {\textstyle{{42} \over 3}} \,\,\,\,n\,\,\rm{odd}\\
 \end{array} \right.
\end{equation}

An exact evaluation of the above costs shows that applying LNN constraints increases CNOT count by at most a factor of $\frac{5}{3}$.

\section{Conclusions and Future Directions} \label{sec:conclusion}
In this paper, a decomposition approach for quantum-logic synthesis was proposed which is based on the previous CSD and QSD methods.
To do this, a basic block based on CSD method was constructed which contains $l$ qubits. Next, the proposed circuit for a generic unitary gate over $l$ qubits was used as a basic block to present our quantum synthesis method. In doing so, the previously proposed QSD method for quantum-logic synthesis was recursively applied to reach unitary gates over $l$ qubits. Then, the proposed basic block was used and further optimizations were applied to remove redundant gates.
To evaluate the proposed method, number of CNOT and one-qubit gates with and without the nearest neighbor constraint was analyzed. According to our analysis, the decomposition level $l$ can be selected to trade off the number of one-qubit gates against CNOT gates. Since producing fewer CNOT gates is more desirable, the decomposition level is set to $\lceil 2n/3 \rceil$ in the proposed decomposition method.

Due to the potential of the proposed decomposition method in consideration of the number of CNOT and one-qubit gates, it is our hope that the proposed synthesis method may lead to further improvement in the number of CNOT and one-qubit gates simultaneously in future.

\section*{Acknowledgment}
We would like to thank Mikko M\"{o}tt\"{o}nen for helpful discussions. 


\end{document}